\documentstyle[12pt]{article}

\begin{document}

\title{Symmetries and Unification }
\author{Yue-Liang Wu \\
Institute of Theoretical Physics \\
       Chinese Academy of Sciences \\
      Beijing, 100080, P.R. China \\E-mail: YLWU@ITP.AC.CN}
\date{AS-ITP-97-29, 1997}
\maketitle

\begin{abstract}
 Symmetries concerning the ordinary coordinate spacetime and internal spacetime 
are discussed. 
A possible unification model of electroweak, strong and gravitational 
interactions is briefly described. 
\end{abstract}

  Symmetry has played a crucial role in physics. I first learned the importance 
of symmetries 
in physics was from the lecture talks given by professor C.S. Wu when I was a 
student in the Department of 
Physics at Nanjing University.  The $\beta$ decay experiment by Wu {\it et al} 
\cite{CSWU} and the $\pi$-$\mu$ decay experiment by Garwin, Lederman and 
Weinrich \cite{GLW} and by Friedman and Telegdi\cite{FT} were the most excellent 
experiments that first decisively established parity nonconservation discovered 
first by T.D. Lee and C.N. Yang \cite{LY}. In fact, these experiments proved 
not only parity $P$ violation, but also charge asymmetry under 
particle-antiparticle 
conjugation $C$.  Late on, the experiment by Christenson, J. Cronin, V.L. Fitch 
and 
R. Turlay \cite{CP} in 1964 established $CP$ violation in kaon decays. From CPT 
symmetry, 
CP violation implies time reversal $T$ asymmetry. All physical 
laws are related to these discrete symmetries as they are the basic symmetries 
of spacetime. 
The present status of these discrete symmetries has been known that: weak 
interaction violates the 
parity $P$, charge conjugation $C$ and $CP$ symmetries, while strong and 
electromagnetic interactions appear to be invariant under these discrete 
symmetries. 

  The papers by  Glashow, Weinberg  and Salam unified the weak interaction 
with electromagnetic force\cite{EW}. Such a unified interaction is called 
electroweak 
interaction characterized by the local gauge symmetry SU(2)$\times$ U(1). The 
strong interaction 
is described by QCD with gauge symmetry group SU(3). The model with gauge 
symmetry group U(1)$\times$ SU(2)$\times$ SU(3) is usually called as the 
standard model which has successfully described the electroweak and strong 
interactions. The gravitational force is characterized by Einstein's general 
relativity which is invariant under general coordinate transformations. 

 However, in the standard model, one has to introduce eighteen parameters to 
describe real 
world. Three charged lepton masses, six quark masses, three quark mixing angles 
and one CP-violating phase, three gauge coupling constants and two weak gauge 
boson masses, all of 
unknown origins. Therefore, the standard model cannot be considered as a 
complete model. 
Thus the outstanding puzzles that confront us today are:

  1) Origin of CP violation and fermion masses.

  2) Basic symmetries of nature and symmetry breaking mechanism.
 
 Numerious efforts have been made to understand these puzzles.  In this talk, 
I briefly outline the progresses in our recent studies. 

   I. {\bf Mechanisms of CP Violation in General 2HDM}. In order to understand 
origin and mechanism of CP violation, we consider a simple extention of the 
standard model by just adding a Higgs doublet. Since the discovery 
of CP violation in 1964, two of the interesting ideas about 
CP-violating scheme have been known as the superweak interaction proposed by 
Wolfenstein\cite{SW} and spontaneous breaking of CP symmetry suggested by 
T.D. Lee\cite{TDL}. These two ideas were found to be simultaneously realized 
in two-Higgs doublet models\cite{2HDM1,2HDM2,WW1}. In the most general two-Higgs 
doublet model 
proposed recently by Wolfenstein and myself\cite{WW1}, we observed that there 
exist rich 
soures of CP violation. In this general 2HDM, we have assumed that CP violation 
arises solely through the Higgs potential and there is no discrete symmetry that 
distingushes the two Higgs bosons. It was found that an approximate global 
family 
symmetry is sufficient to suppress flavor-changing neutral scalar interactions.  
The model have four major sources of CP violation induced from a single 
CP-violating 
phase, i.e., the relative phase of the two vacuum expectation values:

 (A)\  The CKM CP-violating source. Its effects are related to the magnitudes of 
the  
flavor-changing scalar interactions. This is easily understood because in the 
case of 
natural flavor conservation ensured by imposing discrete symmetry, the model is 
CP-conserved.  
In addition to the usual CP violation in $W^{\pm}$ exchange, there is also in 
all two-Higgs doublet models a similar CP violation in the charged-Higgs-boson 
sector.  
    
 (B)\   New CP-violating sources that are independent of the CKM phase. Their 
effects are 
related to the diagonal couplings of the two Higgs doublets to the fermions.  
Therefore, this type of CP violation occurs not only in the charged-Higgs-boson 
exchange 
processes but also in flavor-conserving scalar interactions.

   Let us illustrate the origin of these sources. For this purpose, one can 
simply neglect 
the off-diagonal elements of the couplings. For each fermions, we have 
\begin{equation}
m_{f_{i}} e^{i\delta_{f_{i}}} = (g_{1f_{i}}e^{i\delta} \cos\beta + 
g_{2f_{i}}\sin\beta )v
\end{equation}
where $g_{1f_{i}}$ and $g_{2f_{i}}$ are Yukawa couplings corresponding to the 
two Higgs doublets $\phi_{1}$ and $\phi_{2}$. $\delta$ is the relative phase of 
the two vacuum expectation 
values $v_{1}e^{i\delta}$ and $v_{2}$. The angle $\beta$ is given by $\tan\beta 
= v_{1}/v_{2}$ 
with $\sqrt{v_{1}^{2} + v_{2}^{2}} = v =246$GeV. $m_{i}$ are the fermion masses 
and $\delta_{i}$ are phases associated with the masses. One gets rid of 
$\delta_{i}$ by redefining the corresponding right-handed fermions. These phases 
then enter into the scalar boson interactions with effective couplings
\begin{equation}
(g_{1f_{i}}e^{i\delta} \sin\beta - g_{2f_{i}}\cos\beta )e^{-i\delta_{f_{i}}} 
\equiv 
\xi_{f_{i}}m_{f_{i}}/v
\end{equation} 

 (C)\  Superweak type CP violation. The effective CP-violating phases arise from 
the small off-diagonal couplings of two Higgs doublets. These yield CP violation 
in flavor-changing processes mediated by the exchange of neutral scalar bosons 
(FCNE).

 (D)\  CP violation due to neutral scalar boson mixings. This source arises from 
the matrix 
$O^{H}$ that diagonalizes the Higgs boson mass matrix. Even in the absence of 
fermions this 
$O^{H}$ may violate CP invariance.    

  As a consequence, we observed the following interesting features arising from 
the 
new sources:

 (1) The new sources through charged Higgs boson exchange can make a 
contribution to direct CP-violating parameter $\epsilon'/\epsilon$ which has the 
order of magnitude 
\begin{eqnarray}
\frac{\epsilon'}{\epsilon} & \simeq & 10^{-4} \sim 10^{-5}, \qquad \tan\beta 
\sim 1, \nonumber \\
\frac{\epsilon'}{\epsilon} & \simeq & 10^{-3} , \qquad \tan\beta \sim 10\ 
[m_{H^{+}}/(200GeV)]
\end{eqnarray}
without conflicting with other constraints.  These predictions are comparable 
with those from the standard CKM CP source. 

 (2) The indirect CP-violating paramter $\epsilon$ can be fitted by the new 
sources from box diagrams containing $H^{\pm}$. It may also receive significant 
contribution from superweak 
FCNE.

 (3) CP asymmetry in the decay $b\rightarrow s\gamma $ may arise from the new 
sources from the 
charged Higgs boson interactions with fermions. The asymmetry may be larger than 
in the 
standard model and can lie between $0.01$ and $0.1$ \cite{WW2}. 

 (4)  The new sources may also seriously change the expectations for CP 
violation in the $B^{0}$ 
system. In general, if there are large contributions to $B^{0}-\bar{B}^{0}$ 
mixing from superweak and new sources,  the measured three angles corresponding 
to the unitarity triangle of the CKM matrix may not be closed.

 (5) The new sources could provide large CP violation in hyperon decays. The 
resulting values can reach 
the present experimental sensitivity\cite{CPHyper}. 

 (6) The new sources may also lead to a significant time reversal $T$ violation. 
Such as 
the electric dipole moment $D_{e}$ of the electron and the electric dipole 
moment $D_{n}$ of the neutron. From both charged and neutral Higgs boson 
contributions to $D_{n}$ and $D_{e}$ via the two-loop Barr-Zee mechanism, 
resulting values of $D_{n}$ of the order $10^{-25}$ to $10^{-26}$ {\it e cm} and 
of $D_{e}$ of the order $10^{-26}$ to $10^{-27}$ {\it e cm} close to the present 
limits are allowed without conflicting with other constraints.

  In a word, this general 2HDM, as one of the most simplest extentions of the 
standard model, 
does contain rich physical phenomena. Though more unkown parameters have been 
introduced, 
it does show us from where one may look for the possible new physics.   
 
   II. {\bf Predictive SUSY GUTs}. Let us extend the standard model along the 
direction of supersymmetric grand 
unification theories (SUSY GUTs) for the purposes of reducing the 
parameters. As the eighteen 
parameters in the standard model have been improved to be more and more 
accuracy.  
It reminds us that we are in a stage similar to that of atomic spectroscopy 
before Balmer. Much effort has been made along this direction\cite{RABY}. 
Here I briefly describe an SUSY GUT model proposed recently by 
Chou and myself\cite{CW1,CW2,CW3}. Our SUSY GUT model was based on the symmetry 
group SUSY 
SO(10)$\times \Delta(48)\times$ U(1). Where SO(10)\cite{SO10} unifies all 
leptons and quarks of a single generation into a single 16-dimensional spinor 
representation of SO(10). The dihedral group $\Delta(48)$, a subgroup of SU(3), 
is taken as the family group. U(1) is family-independent and
is introduced to distinguish various fields which belong to the same 
representations of 
SO(10)$\times \Delta(48)$. The irreducible representations of $\Delta(48)$ 
consisting of five 
triplets and three singlets have been found to be sufficient to build an 
interesting texture 
structure for fermion mass matrices. The symmetry $\Delta(48)\times$ U(1) 
naturally ensures 
the texture structure with zeros for Yukawa coupling matrices. To reduce the 
possible free parameters, the universality of coupling constants in the 
superpotential is assumed, i.e., all
the coupling coefficients are assumed to be equal and have the same origins from 
perhaps a more fundamental theory. With these considerations, Yukawa coupling 
matrices which determine the 
masses and mixings of all quarks and leptons can be obtained by carefully 
choosing the 
structure of the physical vacuum and integrating out the heavy fermions at the 
GUT scale
\begin{equation}
\Gamma_{u}^{G} = \frac{2}{3}\lambda_{H}  \left(\begin{array}{ccc} 
0  &  \frac{3}{2}z'_{u} \epsilon_{P}^{2} &   0   \\
\frac{3}{2}z_{u} \epsilon_{P}^{2} &  - 3 y_{u} \epsilon_{G}^{2} e^{i\phi}  & 
-\frac{\sqrt{3}}{2}x_{u}\epsilon_{G}^{2}  \\
0  &  - \frac{\sqrt{3}}{2}x_{u}\epsilon_{G}^{2}  &  w_{u} 
\end{array} \right)
\end{equation}   
and
\begin{equation}
 \Gamma_{f}^{G} = \frac{2}{3}\lambda_{H} \frac{(-1)^{n+1}}{3^{n}} 
\left( \begin{array}{ccc} 
0  &  -\frac{3}{2}z'_{f} \epsilon_{P}^{2} &   0   \\
-\frac{3}{2}z_{f} \epsilon_{P}^{2} &  3 y_{f} 
\epsilon_{G}^{2} e^{i\phi}  
& -\frac{1}{2}x_{f}\epsilon_{G}^{2}  \\
0  &  -\frac{1}{2}x_{f}\epsilon_{G}^{2}  &  w_{f} 
\end{array} \right)
\end{equation}   
for $f=d,e$,  and 
\begin{equation}
\Gamma_{\nu}^{G} = \frac{2}{3}\lambda_{H}\frac{(-1)^{n+1}}{3^n}
\frac{1}{5^{n+1}} 
\left( \begin{array}{ccc} 
0  &  -\frac{15}{2}z'_{\nu} \epsilon_{P}^{2} &   0   \\
-\frac{15}{2}z_{\nu} \epsilon_{P}^{2} &  15 y_{\nu} 
\epsilon_{G}^{2} e^{i\phi}  
& -\frac{1}{2}x_{\nu}\epsilon_{G}^{2}  \\
0  &  -\frac{1}{2}x_{\nu}\epsilon_{G}^{2}  &  w_{\nu} 
\end{array} \right)
\end{equation}   
for Dirac-type neutrino coupling. The Majorana neutrino mass matrix is chosen to 
be 
\begin{equation}
M_{N}^{G} = M_{R} \left( \begin{array}{ccc} 
0  &  0 &   \frac{1}{2}z_{N}\epsilon_{P}^{2} e^{i(\delta_{\nu} + \phi_{3})}   \\
0  &  y_{N} e^{2i\phi_{2}} & 0 \\
\frac{1}{2}z_{N}\epsilon^{2}_{P} 
e^{i(\delta_{\nu} + \phi_{3})}   & 0 &  w_{N}\epsilon_{P}^{4} e^{2i\phi_{3}} 
\end{array} \right)
\end{equation}   

We choose $n=4$ for a realistic case.
$\lambda_{H}=\lambda_{H}^{0}r_{3}= 2\lambda_{t}^{G}/3 $, $\epsilon_{G}\equiv 
(\frac{v_{5}}{v_{10}})
\sqrt{\frac{r_{2}}{r_{3}}}$ and $\epsilon_{P}\equiv
(\frac{v_{5}}{\bar{M}_{P}})\sqrt{\frac{r_{1}}{r_{3}}}$ are three parameters. 
Where $\lambda_{H}^{0}$ is a universal 
coupling constant expected to be of order one, $r_{1}$, $r_{2}$ and $r_{3}$ 
denote the ratios of the coupling constants of the superpotential at 
the GUT scale for the textures `12', `22' (`32') and `33' respectively. 
They represent the possible renormalization group (RG) effects 
running from the scale $\bar{M}_{P}$ to the GUT scale. 
$\bar{M}_{P}$, $v_{10}$ and $v_{5}$ are the VEVs for 
U(1)$\times \Delta(48)$, SO(10) and SU(5) symmetry breaking
respectively. $\phi$ is the physical CP phase\footnote{ We have rotated
away other possible phases by a phase redefinition of the fermion fields.} 
arising from the VEVs. The assumption of maximum CP violation implies that 
$\phi = \pi/2$. $M_{R} = \lambda_{H}\epsilon_{P}^{4} 
\epsilon_{G}^{2}v_{10}^{2}/\bar{M}_{P}$, $\lambda_{1}^{N} =
\epsilon_{P}^{2}M_{R}$, $\lambda_{2}^{N} = M_{R}/\epsilon_{G}^{2}$ and 
$\lambda_{3}^{N} = \epsilon_{P}^{4}M_{R}$.  
$x_{f}$, $y_{f}$, $z_{f}$, and $w_{f}$ $(f = u, d, e, \nu)$ and $y_{N}$, $z_{N}$ 
and $w_{N}$
are the Clebsch factors of SO(10) determined by the 
directions of symmetry breaking of the adjoints {\bf 45}'s. 
The resulting Clebsch  factors are
$ w_{u}=w_{d}=w_{e}=w_{\nu} =1$, $ x_{u}= 5/9,\  x_{d}= 7/27,\  x_{e}=-1/3,\  
x_{\nu} = 1/5$, $y_{u}=0, \  y_{d}=y_{e}/3=2/27, \ y_{\nu} = 4/225$, $z_{u}=1, \ 
 z_{d}=z_{e}= -27,\   z_{\nu} = -15^3 = -3375$, 
$ z'_u = 1-5/9 = 4/9,\  z'_d = z_d + 7/729 \simeq z_{d}$,
$z'_{e} = z_{e} - 1/81 \simeq z_{e},\   
z'_{\nu} = z_{\nu} + 1/15^{3} \simeq z_{\nu}$.   
$y_{N} = 9/25, \  z_{N}= 4, \ w_{N} = 256/27 $

 By diagonalizing the mass matrices and taking into account the renormalization 
group effects from GUT scale down to low energies, 
we can obtain twenty three predictions with four input parameters. Four  
predictions for $|V_{us}|$, $|V_{ub}/V_{cb}|$, $|V_{ub}/V_{cb}|$ and
$m_{s}/m_{d}$ are RG scaling-independent. All the predictions are presented in 
Table 1. 

\newpage 

Table 1. Output observables and model parameters and their 
predicted values with input parameters  $m_{e}$ = 0.511 MeV, 
$m_{\mu}$ = 105.66 MeV, $m_{\tau}$ = 1.777 GeV and $m_{b}(m_{b})$ = 4.32 GeV for 
$\alpha_{s}(M_{Z})$ = 0.113
\\
\begin{tabular}{|c|c|c|c|c|}   \hline
 Output para.   &  Values    &  Data  & 
 Output para.   &  Values    \\ \hline 
$M_{t}$\ [GeV]  &  179   &  $175\pm 6 $  &  $J_{CP} = A^{2} 
\lambda^{6} \eta $ & $2.62 \cdot 10^{-5}$  \\
$m_{c}(m_{c})$\ [GeV]  &  1.21   & $1.27 \pm 0.05$  & 
 $\alpha$ & $86.28^{\circ}$ \\ 
$m_{u}$(1GeV)\ [MeV]  &  4.11   &  $4.75 \pm 1.65$ & 
$\beta$ & $22.11^{\circ}$ \\
$m_{s}$(1GeV)\ [MeV]  &  156.5  &  $165\pm 65$  &  
$\gamma$ & $71.61^{\circ}$  \\
$m_{d}$(1GeV) \ [MeV]  &  6.26 & $8.5 \pm 3.0$ & 
$m_{\nu_{\tau}}$ [eV]  & $ 2.4515$    \\
$|V_{us}|=\lambda $ & 0.22 & $0.221 \pm 0.003$ & 
$m_{\nu_{\mu}}$ [eV]  & $2.4485$   \\
$\frac{|V_{ub}|}{|V_{cb}|} $ & 0.083 & 
$0.08 \pm 0.03$ & $m_{\nu_{e}}$ [eV] & $ 1.27\cdot 10^{-3}$   \\
$\frac{|V_{td}|}{|V_{ts}|} $ & 0.209 & 
$0.24 \pm 0.11$ & $m_{\nu_{s}}$ [eV]  & $ 2.8 \cdot 10^{-3}$  \\
 $|V_{cb}|=A\lambda^{2}$ & 0.0389  &  $0.039 \pm 0.005 $ &
  $|V_{\nu_{\mu}e}| $ &  -0.049  \\
$\lambda_{t}^{G}$  & 1.20  & - &  $|V_{\nu_{e}\tau}| $ &  0.000  \\
$\tan \beta = v_{2}/v_{1}$ & 2.12 & - &   $|V_{\nu_{\tau}e}| $ & -0.049   \\
$\epsilon_{G}$ &  $0.2987$ & - & 
$|V_{\nu_{\mu}\tau}| $ &  -0.707  \\
$\epsilon_{P}$  & $0.0101 $ & - & 
$|V_{\nu_{e}s}|$ & $ 0.038 $ \\
$B_{K}$ & 0.96 &  $0.82 \pm 0.10$ & $M_{N_{1}}$ [GeV] & $\sim 361$  \\
$f_{B}\sqrt{B}$ [MeV] & 212  & $200 \pm 70 $ &
$M_{N_{2}}$ [GeV] & $1.77\cdot 10^{6}$  \\
 Re($\varepsilon'/\varepsilon)/10^{-3} $ & $1.4 \pm 1.0 $ &  
$1.5 \pm 0.8 $ & $M_{N_{3}}$ [GeV] & $ 361$  
  \\  \hline
\end{tabular}
\\

From the above results, we observe the following features for the neutrinos

1. a $\nu_{\mu}(\bar{\nu}_{\mu}) \rightarrow \nu_{e} (\bar{\nu_{e}})$ 
short wave-length oscillation with 
\begin{equation}
\Delta m_{e\mu}^{2} = m_{\nu_{\mu}}^{2} - m_{\nu_{e}}^{2} 
\simeq 6\  eV^{2}, \qquad
\sin^{2}2\theta_{e\mu} \simeq 1.0 \times 10^{-2} \ , 
\end{equation}
which is consistent with the LSND experiment\cite{LSND} 
\begin{equation}
\Delta m_{e\mu}^{2} = m_{\nu_{\mu}}^{2} - m_{\nu_{e}}^{2} 
\simeq (4-6) eV^{2}\ , \qquad
\sin^{2}2\theta_{e\mu} \simeq 1.8 \times 10^{-2} \sim 3 \times 10^{-3}\ ; 
\end{equation}
 
2. a $\nu_{\mu} (\bar{\nu}_{\mu}) \rightarrow \nu_{\tau} (\bar{\nu}_{\tau})$
long-wave length oscillation with 
\begin{equation}
\Delta m_{\mu\tau}^{2} = m_{\nu_{\tau}}^{2} - m_{\nu_{\mu}}^{2} \simeq 
1.5 \times 10^{-2} eV^{2}\ , \qquad
\sin^{2}2\theta_{\mu\tau} \simeq 0.987 \ ,
\end{equation}
which could explain the atmospheric neutrino 
deficit\cite{ATMO}:
\begin{equation}
\Delta m_{\mu\tau}^{2} = m_{\nu_{\tau}}^{2} - m_{\nu_{\mu}}^{2} \simeq 
(0.5-2.4) \times 10^{-2} eV^{2}\ , \qquad
\sin^{2}2\theta_{\mu\tau} \simeq 0.6 - 1.0 \ ,
\end{equation}
with the best fit\cite{ATMO}
\begin{equation}
\Delta m_{\mu\tau}^{2} = m_{\nu_{\tau}}^{2} - m_{\nu_{\mu}}^{2} \simeq 
1.6\times  10^{-2} eV^{2}\ , \qquad
\sin^{2}2\theta_{\mu\tau} \simeq 1.0 \ ;
\end{equation}

3. Two massive neutrinos $\nu_{\mu}$ and $\nu_{\tau}$ with 
\begin{equation}
m_{\nu_{\mu}} \simeq m_{\nu_{\tau}}  \simeq 2.45\  eV \ ,
\end{equation}
 fall in the range required by  
possible hot dark matter\cite{DARK}.

4. $(\nu_{\mu} - \nu_{\tau})$ oscillation will be beyond the reach
of CHORUS/NOMAD and E803. However, $(\nu_{e} - \nu_{\tau})$ oscillation
may become interesting as a short wave-length oscillation with 
\begin{equation}
\Delta m_{e\tau}^{2} = m_{\nu_{\tau}}^{2} - m_{\nu_{e}}^{2} 
\simeq 6\  eV^{2}, \qquad
\sin^{2}2\theta_{e\tau} \simeq 1.0 \times 10^{-2} \ , 
\end{equation}
which should provide an independent test on the pattern of the present 
Majorana neutrino mass matrix. 
 
5. Majorana neutrino  allows neutrinoless double beta decay
$(\beta \beta_{0\nu})$\cite{DBETA}. Its decay amplitude is known to
depend on the masses of
Majorana neutrinos $m_{\nu_{i}}$ and the lepton mixing 
matrix elements $V_{ei}$. 
The present model is compatible with the present experimental upper bound
on neutrinoless double beta decay
\begin{equation} 
\bar{m}_{\nu_{e}} = \sum_{i=1}^{3} [ V_{ei}^{2} m_{\nu_{i}} \zeta_{i} ] 
\simeq 1.18 \times 10^{-2}\ eV\ < \  \bar{m}_{\nu}^{upper} \simeq 0.7 \ eV  
\end{equation}  
The decay rate is found to be
\begin{equation}
\Gamma_{\beta\beta} \simeq \frac{Q^{5}G_{F}^{4}\bar{m}_{\nu_{e}}^{2}
p_{F}^{2}}{60\pi^{3}} \simeq 1.0 \times 10^{-61} GeV
\end{equation}
with the two electron energy $Q\simeq 2$ MeV and $p_{F}\simeq 50$ MeV. 

6.  In this case, solar neutrino deficit has to be explained by oscillation 
between $\nu_{e}$ and  a sterile neutrino $\nu_{s}$
\cite{STERILE}. Since strong bounds on the number of neutrino species both from 
the invisible $Z^{0}$-width and from primordial nucleosynthesis \cite{NS} 
require the 
additional neutrino to be sterile (singlet of SU(2)$\times$ U(1), or 
singlet of SO(10) in the GUT SO(10) model). 
Masses and mixings of the triplet sterile neutrinos can be chosen 
by introducing an additional  singlet scalar with VEV $v_{s}\simeq 336$ GeV.
We find  
\begin{eqnarray}
& & m_{\nu_{s}} = \lambda_{H} v_{s}^{2}/v_{10} \simeq 2.8 \times 10^{-3} eV
\nonumber \\
& & \sin\theta_{es} \simeq \frac{m_{\nu_{L}\nu_{s}}}{m_{\nu_{s}}} 
= \frac{v_{2}}{2v_{s}} \frac{\epsilon_{P}}{\epsilon_{G}^{2}} \simeq 3.8 
\times 10^{-2} 
\end{eqnarray}
with the mixing angle  consistent with the requirement necessary for 
primordial nucleosynthesis \cite{PNS} 
given in \cite{NS}.  The resulting parameters
\begin{equation}
\Delta m_{es}^{2} = m_{\nu_{s}}^{2} - m_{\nu_{e}}^{2} \simeq 6.2 \times 
10^{-6} eV^{2}, \qquad  \sin^{2}2 \theta_{es} \simeq 5.8 \times 10^{-3}
\end{equation}
are consistent with the values \cite{STERILE} obtained from 
fitting the experimental data:
\begin{equation}
\Delta m_{es}^{2} = m_{\nu_{s}}^{2} - m_{\nu_{e}}^{2} \simeq (4-9) \times 
10^{-6} eV^{2}, \qquad  \sin^{2}2 \theta_{es} \simeq (1.6-14) \times 10^{-3}
\end{equation}

  This scenario can be tested by the next generation solar neutrino 
experiments in Sudhuray Neutrino Observatory (SNO) and 
Super-kamiokanda (Super-K), both planning to start 
operation in 1996. From measuring  neutral current events, one could 
identify $\nu_{e} \rightarrow \nu_{s}$ or 
$\nu_{e} \rightarrow \nu_{\mu} (\nu_{\tau})$ since the sterile 
neutrinos have no weak gauge interactions. From measuring seasonal
variation, one can further distinguish the small-angle MSW \cite{MSW} 
oscillation from vacuum mixing oscillation. 

  III. {\bf Unification of All Basic Forces}. We finally consider a possible 
unification 
of the standard model with gravity. This is one of the great theoretical 
endeavours in this century. One of the difficulties arises from the no-go 
theorem.  Most of the attempts to unify all basic forces involve higher 
dimensional spacetime, such as Kaluza-Klein Yang-Mills theories, supergravity 
theories and superstring theories, {\it etc}. The Kaluza-Klein approach is not 
rich enough to support the fermionic representations of the standard model. The 
maximum supergravity has SO(8) symmetry
which is too small to include the standard model. In superstring theories, all 
the known particle interactions can be reproduced, but millions of vacua have 
been found. The outstanding problem is to find which one is the true vacuum of 
the theory. We then presented an alternative scheme\cite{CW4}. Firstly, we 
observe that quarks and leptons in the standard model can be 
unified into a single 16-dimensional representation of complex chiral spinors in 
SO(10)\cite{SO10}. Each complex chiral spinor belong to a single 4-dimensional 
representation of SO(1,3). In an unified theory, it is an attractive idea to 
treat these 64 real spinor components on the same footing, i.e., they have to be 
a single representation of a larger group. It is therefore natural to consider 
SO(1,13) as our unified group and the gauge potential of SO(1,13) as the 
fundamental interaction that unifies the four basic forces (strong, 
electromagnetic, weak and gravitational) of nature. Secondly, to avoid the 
restriction given by no-go theorem and other problems mentioned above, we 
consider the ordinary coordinate spacetime remains to be a 4-dimensional 
manifold $S_{4}$ with metric $g_{\mu\nu}(x)$, $\mu$, $\nu$=0,1,2,3. At each 
point P: $x^{\mu}$, there is an d-dimensional flat space $M_{d}$ with $d > 4$ 
and signature (1, -1, $\cdots$, -1). We assume that the tangent space $T_{4}$ of 
$S_{4}$ at point P to be an 4-dimensional submanifold of $M_{d}$ spanned by four 
vectors $e_{\mu}^{A}(x)$ $\mu$=0,1,2,3; 
$A\equiv (\alpha, a)$ with $\alpha=0, 1, 2, 3$ and $a=1, \cdots, d-4$  such that 
\begin{equation}
g_{\mu\nu}(x) = e_{\mu}^{A}(x)e_{\nu}^{A}(x)\eta_{AB}
\end{equation}
where $\eta_{AB} = diag.(1, -1, \cdots, -1)$ can be considered as the metric of 
the flat space $M_{d}$. We shall call $e_{\mu}^{A}(x)$ to be the generalized 
vierbein fields or simply the frame fields. 

Under general coordinate transformations and the rotations in $M_{d}$, 
$e_{\mu}^{A}(x)$ transform as a covariant vector in ordinary coordinate spactime 
and a vector in the $M_{d}$ rotation, $e_{m}^{A}(x)$ transform as a covariant 
vector in the $C_{d-4}$ rotation and a vector in the $M_{d}$ rotation. For a 
theory to be invariant under both general coordinate transformations and local 
rotations in the flate space $M_{d}$, it is necessary to introduce affine 
connection $\Gamma_{\mu\nu}^{\rho}(x)$ for general coordinate transformations 
and gauge potential $\Omega_{\mu}^{AB}(x)= - \Omega_{\mu}^{BA}(x)$ for 
d-dimensional rotation SO(1,d-1) in $M_{d}$. These transformations are connected 
by the requirement that $T_{4}$ has to be the submanifold of $M_{d}$ spanned by 
four vectors $e_{\mu}^{A}(x)$ at point P and $e_{\mu}^{A}(x)$ should be a 
covariantly constant frame and satisfy the condition
\begin{equation} 
D_{\mu}e_{\rho}^{A} = \partial_{\mu}e_{\rho}^{A} -\Gamma_{\mu 
\rho}^{\sigma}e_{\sigma}^{A} + g_{U}\Omega_{\mu B}^{A}e_{\rho}^{B} = 0
\end{equation}
It is then easily verified that 
\begin{eqnarray}
D_{\mu}g_{\rho\sigma} & = & \partial_{\mu} g_{\rho\sigma} - 
\Gamma_{\mu\rho}^{\lambda}g_{\lambda\sigma} - 
\Gamma_{\mu\sigma}^{\lambda}g_{\rho\lambda}= 0  \\
D_{\mu}e^{\rho}_{A} & = & \partial_{\mu}e^{\rho}_{A} + \Gamma_{\mu 
\sigma}^{\rho}e^{\sigma}_{A} - g_{U}\Omega_{\mu A}^{B}e^{\rho}_{B} = 0 
\end{eqnarray}

 With the above considerations, we can now construct an invariant action under 
general coordinate transformations in the ordinary coordinate spacetime and the 
local  SO(1,d-1) group symmetry in $M_{d}$ with $D_{\mu}e_{\nu}^{A}=0$ as a 
constraint. In addition, the action is required to have no dimensional 
parameters and to be renormalizable in the sense of the power counting. The 
general form of the action which satisfies these requirements is  
\begin{eqnarray}
S_{B} & = & \int d^{4}x \sqrt{-g} \{   
- \frac{1}{4} 
F_{\mu\nu}^{AB}F_{\rho\sigma}^{CD}g^{\mu\rho}g^{\nu\sigma}\eta_{AC}\eta_{BD} 
\nonumber \\  
& - & \frac{1}{2}\xi \phi^{2} F_{\mu\nu}^{AB} e^{\mu}_{A}e^{\nu}_{B} 
+ \frac{1}{2}g^{\mu\nu}\partial_{\mu}\phi \partial_{\nu}\phi  + 
\frac{1}{4}\lambda \phi^{4} \\
& + & \zeta F_{\mu\nu}^{AB}F_{\rho\sigma}^{CD}g^{\mu\rho}\eta_{AC} 
e^{\nu}_{B}e^{\sigma}_{D} 
+ a_{1}F_{\mu\nu}^{AB}F_{\rho\sigma}^{CD} 
e^{\mu}_{C}e^{\nu}_{D}e^{\rho}_{A}e^{\sigma}_{B} \nonumber \\
& + & a_{2}F_{\mu\nu}^{AB}F_{\rho\sigma}^{CD}e^{\mu}_{C}
e^{\nu}_{B}e^{\rho}_{A}e^{\sigma}_{D} + a_{3} F_{\mu\nu}^{AB}F_{\rho\sigma}^{CD}
e^{\mu}_{A}e^{\nu}_{B}e^{\rho}_{C}e^{\sigma}_{D} \} \nonumber 
\end{eqnarray}
where $\phi(x)$ is a scalar field introduced to avoid the dimensional coupling 
constants. $a_{i}$ (i=1,2,3), $\zeta$, $\xi$ and $\lambda$ are dimensionless 
parameters. $F_{\mu\nu}^{AB}$ is the field strength defined in a standard way 
\begin{equation}
F_{\mu\nu}^{AB} = \partial_{\mu}\Omega_{\nu}^{AB} - 
\partial_{\nu}\Omega_{\mu}^{AB} + g_{U}(\Omega_{\mu C}^{A} \Omega_{\nu}^{CB} - 
\Omega_{\nu C}^{A} \Omega_{\mu}^{CB})
\end{equation}
The tensor $F_{\mu}^{A}$ is defined as $F_{\mu}^{A} = 
F_{\mu\nu}^{AB}e^{\nu}_{B}$

   Note that not all the gauge fields $\Omega_{\mu}^{AB}(x)$ are simply new 
propagating fields 
due to the constraints $D_{\mu}e_{\rho}^{A}=0$. By counting the constraint 
equations ($4\times 4 \times$d), unknowns $\Omega_{\mu}^{AB}(x)$ (with 4d(d-1)/2 
degrees of freedom) and $e_{\mu}^{A}(x)$ (with $4\times$d degrees of freedom) as 
well as $\Gamma_{\mu\sigma}^{\rho}$ (with 40 degrees of freedom for the 
symmetric parts $\Gamma_{(\mu\sigma)}^{\rho}= \Gamma_{(\sigma\mu)}^{\rho}$ and 
24 degrees of freedom for antisymmetric parts $\Gamma_{[\mu\sigma]}^{\rho}= - 
\Gamma_{[\sigma\mu]}^{\rho}$ ), one sees that besides the antisymmetric parts 
$\Gamma_{[\mu\sigma]}^{\rho}$, the independent degrees of freedom are (4d + 
4(d-4)(d-5)/2). These independent degrees of freedom coincide with the degrees 
of freedom of the frame fields $e_{\mu}^{A}(x)$ and the gauge fields 
$A_{\mu}^{ab}(x)$ ($a,b = 1, \cdots, d-4$) of the subgroup SO(d-4). In addition, 
the gauge conditions in the coset SO(1,d-1)/SO(d-4) lead to additional 
constraints (4d-10). Thus the independent degrees of freedom are reduced to (10 
+ 4(d-4)(d-5)/2) which exactly match with the degrees of freedom of the metric 
tensor $g_{\mu\nu}(x)$ and the gauge fields $A_{\mu}^{ab}(x)$ of the group 
SO(d-4). For d=14, the resulting independent degrees of freedom of the fields 
are sufficient to describe the four basic forces. Where the general relativity 
of the Einstein theory is described by the metric tensor. Photon, W-bosons and 
gluons, that mediate the electromagnetic, weak and strong interactions 
respectively, are different manifestations of the gauge potential 
$A_{\mu}^{ab}(x)$ of the symmetry group SO(10)\cite{SO10}. The curvature tensor 
$R_{\mu\nu\sigma}^{\rho}$ and the Ricci tensor 
$R_{\nu\sigma}=R_{\mu\nu\sigma}^{\rho} g_{\rho}^{\mu}$ as well as the scalar 
curvature $R = R_{\nu\sigma}g^{\nu\sigma}$ are simply related to the field 
strength $F_{\mu\nu}^{AB}$ via   $R_{\mu\nu\sigma}^{\rho} = 
g_{U}F_{\mu\nu}^{AB}e^{\rho}_{A}e_{\sigma B}$, $R_{\nu\sigma}= 
g_{U}F_{\mu\nu}^{AB}e^{\mu}_{A}e_{\sigma B}$ and $R = 
g_{U}F_{\mu\nu}^{AB}e^{\mu}_{A}e^{\nu}_{B}$.

   In the real world, there exist three generations of quarks and leptons. Each 
generation of the quarks and leptons have 64 real degrees of freedom. These 
degrees of freedom will be represented by the 64 compotents of a single Weyl 
fermion $\Psi_{+}(x)$ belonging to the fundamental spinor representation of 
SO(1,13). The action for fermions is given by 
\begin{equation}
S_{F}= \int d^{4}x \sqrt{-g} \{ \frac{1}{2}\bar{\Psi}_{+} e^{\mu}_{A}\Gamma^{A}( 
i\partial_{\mu} 
+ g_{U}\Omega_{\mu}^{BC}\frac{1}{2}\Sigma_{BC} ) \Psi_{+} + h.c.\}
\end{equation}
where $\Sigma_{AB}$ are the generators of the SO(1, d-1) in the spinor 
representations and given by $\Sigma_{AB} =\frac{i}{4}[\Gamma_{A}, \Gamma_{B}]$. 
$\Gamma^{A}$ are the gamma matrices that obey $\{\Gamma^{A}, \Gamma^{B}\} = 2 
\eta^{AB}$. 

  It is not difficult to check that the action can be decomposed into two parts. 
One of the parts has the same form as the action of a multiplicatively 
renormalized unified gauge theory including so-called $R^{2}$-gravity and a 
renormalizable scalar matter field as well a nonminimal gravitational-scalar 
coupling. Another part represents the direct interactions between the gauge 
fields and the gravitational fields. It is expected that such a model has 
provided us a new insight for unifying all the basic forces within the framework 
of quantum field theory. Though the ideas and the resulting model are both 
simple, there remains more theoretical work and experimental efforts needed to 
test whether they are the true choice of nature. 

\section*{Acknowledgments}
 The author would like to thank professor F. Wang for his kind hospitality to 
present this talk which was dedicated 
 to the Memory of professor C.S. Wu.  This work is supported in part by 
Outstanding Young Scientist Fund of China. 
   
%\begin{references}

\end{document}